\documentstyle[twoside]{article}

\catcode`\@=11
\long\def\@makefntext#1{ 
\protect\noindent \hbox to 3.2pt {\hskip-.9pt
$^{{\eightrm\@thefnmark}}$\hfil}#1\hfill} 

\def\thefootnote{\fnsymbol{footnote}}
 \def\@makefnmark{\hbox to 0pt{$^{\@thefnmark}$\hss}}  

\def\ps@myheadings{\let\@mkboth\@gobbletwo
\def\@oddhead{\hbox{} 
\rightmark\hfil\eightrm\thepage}
\def\@oddfoot{}\def\@evenhead{\eightrm\thepage\hfil 
\leftmark\hbox{}}\def\@evenfoot{}
\def\sectionmark##1{}\def\subsectionmark##1{}}

\oddsidemargin=\evensidemargin
\renewcommand{\thefootnote}{\fnsymbol{footnote}}

\newcounter{sectionc}\newcounter{subsectionc}\newcounter{subsubsectionc}
\renewcommand{\section}[1] {\vspace{12pt}\addtocounter{sectionc}{1}
\setcounter{subsectionc}{0}\setcounter{subsubsectionc}{0}\noindent
        {\tenbf\thesectionc. #1}\par\vspace{5pt}}
\renewcommand{\subsection}[1] {\vspace{12pt}\addtocounter{subsectionc}{1}
        \setcounter{subsubsectionc}{0}\noindent
        {\bf\thesectionc.\thesubsectionc. {\kern1pt \bfit #1}}\par\vspace{5pt}}
\renewcommand{\subsubsection}[1] {\vspace{12pt}\addtocounter{subsubsectionc}{1}
        \noindent{\tenrm\thesectionc.\thesubsectionc.\thesubsubsectionc.
        {\kern1pt \tenit #1}}\par\vspace{5pt}}
\newcommand{\nonumsection}[1] {\vspace{12pt}\noindent{\tenbf #1}
        \par\vspace{5pt}}

\newcounter{appendixc}
\newcounter{subappendixc}[appendixc]
\newcounter{subsubappendixc}[subappendixc]
\renewcommand{\thesubappendixc}{\Alph{appendixc}.\arabic{subappendixc}}
\renewcommand{\thesubsubappendixc}
        {\Alph{appendixc}.\arabic{subappendixc}.\arabic{subsubappendixc}}

\renewcommand{\appendix}[1] {\vspace{12pt}
        \refstepcounter{appendixc}
        \setcounter{figure}{0}
        \setcounter{table}{0}
        \setcounter{lemma}{0}
        \setcounter{theorem}{0}
        \setcounter{corollary}{0}
        \setcounter{definition}{0}
        \setcounter{equation}{0}
        \renewcommand{\thefigure}{\Alph{appendixc}.\arabic{figure}}
        \renewcommand{\thetable}{\Alph{appendixc}.\arabic{table}}
        \renewcommand{\theappendixc}{\Alph{appendixc}}
        \renewcommand{\thelemma}{\Alph{appendixc}.\arabic{lemma}}
        \renewcommand{\thetheorem}{\Alph{appendixc}.\arabic{theorem}}
        \renewcommand{\thedefinition}{\Alph{appendixc}.\arabic{definition}}
        \renewcommand{\thecorollary}{\Alph{appendixc}.\arabic{corollary}}
        \renewcommand{\theequation}{\Alph{appendixc}.\arabic{equation}}
        \noindent{\tenbf Appendix \theappendixc #1}\par\vspace{5pt}}
\newcommand{\subappendix}[1] {\vspace{12pt}
        \refstepcounter{subappendixc}
        \noindent{\bf Appendix \thesubappendixc. {\kern1pt \bfit #1}}
        \par\vspace{5pt}}
\newcommand{\subsubappendix}[1] {\vspace{12pt}
        \refstepcounter{subsubappendixc}
        \noindent{\rm Appendix \thesubsubappendixc. {\kern1pt \tenit #1}}
        \par\vspace{5pt}}

\newcommand{\textlineskip}{\baselineskip=13pt}
\newcommand{\smalllineskip}{\baselineskip=10pt}

\def\eightcirc{
\begin{picture}(0,0)
\put(4.4,1.8){\circle{6.5}}
\end{picture}}
\def\eightcopyright{\eightcirc\kern2.7pt\hbox{\eightrm c}}

\newcommand{\publisher}[2]{{\begin{center}\eightrm\smalllineskip
        Received #1\\
        Revised #2
        \end{center}
        }}

\def\abstracts#1#2#3{{
        \centering{\begin{minipage}{4.5in}\baselineskip=10pt\eightrm
        \parindent=0pt #1\par
        \parindent=15pt #2\par
        \parindent=15pt #3
        \end{minipage} }\par}}

\def\keywords#1{{
        \centering{\begin{minipage}{4.5in}\baselineskip=10pt\eightrm
        {\eightit Keywords}\/: #1
        \end{minipage} }\par }}

\renewenvironment{thebibliography}[1]                   
        {\ninerm
         \baselineskip=11pt                             
         \begin{list}{\arabic{enumi}.}
        {\usecounter{enumi}\setlength{\parsep}{0pt}
         \setlength{\leftmargin 17pt}{\rightmargin 0pt} 
         \setlength{\itemsep}{0pt} \settowidth          
        {\labelwidth}{#1.}\sloppy}}{\end{list}}

\newcounter{itemlistc}
\newcounter{romanlistc}
\newcounter{alphlistc}
\newcounter{arabiclistc}

\newcommand{\fcaption}[1]{
        \refstepcounter{figure}
        \setbox\@tempboxa = \hbox{\eightrm Fig.~\thefigure. #1}
        \ifdim \wd\@tempboxa > 5in
           {\begin{center}
        \parbox{5in}{\eightrm \smalllineskip Fig.~\thefigure. #1 }
            \end{center}}
        \else
             {\begin{center}
             {\eightrm Fig.~\thefigure. #1}
              \end{center}}
        \fi}

\newcommand{\tcaption}[1]{
        \refstepcounter{table}
        \setbox\@tempboxa = \hbox{\eightrm Table~\thetable. #1}
        \ifdim \wd\@tempboxa > 5in
           {\begin{center}
        \parbox{5in}{\eightrm\smalllineskip Table~\thetable. #1 }
            \end{center}}
        \else
             {\begin{center}
             {\eightrm Table~\thetable. #1}
              \end{center}}
        \fi}

\def\@citex[#1]#2{\if@filesw\immediate\write\@auxout    
        {\string\citation{#2}}\fi                       
\def\@citea{}\@cite{\@for\@citeb:=#2\do                 
        {\@citea\def\@citea{,}\@ifundefined             
        {b@\@citeb}{{\bf ?}\@warning
        {Citation `\@citeb' on page \thepage \space undefined}}
        {\csname b@\@citeb\endcsname}}}{#1}}

\newif\if@cghi
\def\cite{\@cghitrue\@ifnextchar [{\@tempswatrue
        \@citex}{\@tempswafalse\@citex[]}}
\def\citelow{\@cghifalse\@ifnextchar [{\@tempswatrue
        \@citex}{\@tempswafalse\@citex[]}}
\def\@cite#1#2{{$\null^{#1}$\if@tempswa\typeout
        {IJCGA warning: optional citation argument
        ignored: `#2'} \fi}}

\def\pmb#1{\setbox0=\hbox{#1}
        \kern-.025em\copy0\kern-\wd0
        \kern.05em\copy0\kern-\wd0
        \kern-.025em\raise.0433em\box0}

\def\fnt#1#2{\footnotetext{\kern-.3em
        {$^{\mbox{\scriptsize #1}}$}{#2}}}

\def\runninghead#1#2{\pagestyle{myheadings}
\markboth{{\eightit{\quad #1}}\hfill}{\hfill{\eightit{#2\quad}}}}

\font\tenbf=cmbx10
\font\tenit=cmti10
\font\tenit=cmti10
\font\bfit=cmbxti10 at 10pt
 1
 1
 1

\font\ninerm=cmr9

\font\eightrm=cmr8
\font\eightit=cmti8

\font\tenrm=cmr10

\def\qed{\hbox{${\vcenter{\vbox{                          
   \hrule height 0.4pt\hbox{\vrule width 0.4pt height 6pt
   \kern5pt\vrule width 0.4pt}\hrule height 0.4pt}}}$}}

\newcommand{\bce}{\begin{center}}
\newcommand{\ece}{\end{center}}
\newcommand{\bmath}{\begin{math}}
\newcommand{\emath}{\end{math}}
\newcommand{\bp}{\begin{minipage}[t]}
\newcommand{\ep}{\end{minipage}}

\newcommand{\beq}{\begin{equation}}
\newcommand{\eeq}{\end{equation}}

\def\tmpind#1{\par\noindent\hangindent=0.7truecm\hangafter=1}

\runninghead{V.~Buchholtz \& T.~P\"oschel}
{A Vectorized Algorithm for Molecular Dynamics}

\begin{document}


\thispagestyle{empty}
\setcounter{page}{1}

\renewcommand{\thefootnote}{\fnsymbol{footnote}} 
\def\bsc{{\sc a\kern-6.4pt\sc a\kern-6.4pt\sc a}}
\def\bflatex{\bf L\kern-.30em\raise.3ex\hbox{\bsc}\kern-.14em
T\kern-.1667em\lower.7ex\hbox{E}\kern-.125em X}

\title{\vspace*{-1.7cm}
        \smalllineskip{\flushleft
	%
        {\eightrm International Journal of Modern Physics C}\\
        {\eightrm $\eightcopyright$\, World Scientific Publishing Company}
	%
	%
	\\
        \vspace*{2.1cm}
        {\normalsize\bf A VECTORIZED ALGORITHM FOR MOLECULAR DYNAMICS OF SHORT
RANGE INTERACTING PARTICLES}}}

\author{\footnotesize VOLKHARD BUCHHOLTZ\\
        {\normalsize and} \\
        {\footnotesize THORSTEN P\"OSCHEL} \\
        {\small\em Humboldt--Universit\"at zu Berlin, Institut
 f\"ur Theoretische Physik,}\\
        {\small\em Unter den Linden 6, D--10099 Berlin, Germany}
        }

\date{ }

\maketitle

\vspace{0.10truein}
\publisher{(received date)}{(revised date)}

\vspace*{0.15truein}

\abstracts{
\noindent
We report on a lattice based algorithm, completely vectorized for
molecular dynamics simulations. Its algorithmic complexity is of the
order
$O(N)$, where $N$ is the number of particles. The algorithm works very
effectively
when the particles have short range interaction, but it is applicable
to each kind of interaction. The code was tested on a {\sc Cray ymp el}
in a simulation of flowing granular material.
}{}{}

\vspace*{0.15truein}
\vspace*{0.10truein}


%

\vspace*{1pt}\textlineskip
\section{Introduction}
\vspace*{-0.5pt}
\setcounter{topnumber}{3}
\setcounter{bottomnumber}{3}
\setcounter{totalnumber}{5}
\renewcommand{\topfraction}{0.8}
\renewcommand{\bottomfraction}{0.8}
\noindent
The properties of systems consisting of many interacting particles
have been subjects of interest to scientists for hundreds of years.
Since the equations of motion do not allow for an analytical
solution\cite{arnold} one has to calculate the trajectories of the particles by
integrating {\sc Newton}s equation of motion
numerically for each particle $i$:
\par
\begin{equation}
\ddot{\vec{r}_i} = \frac{1}{m_i} \cdot \vec{F}_{i}(\vec{r}_j) \hspace{0.5in}
(j = 1,\ldots ,N)
\end{equation}
\par
\noindent
This method is called molecular dynamics simulation.
Molecular dynamics simulations have led to many important results,
especially in fluid
dynamics, solid state physics, polymer physics, and plasma physics.
\par
There are many methods to integrate sets of ordinary differen\-tial
equations.\cite{recipes} One of the most common is the
{\sc Gear}--predictor--corrector method\cite{predictor} which we used in our
simulations (see Appendix).
\par
The main problem in molecular dynamics is the calculation
of the forces acting upon each of the particles.  The
term $\vec{F}_{i}(\vec{r}_j)$ may be very difficult to determine, but even if
it is
not too difficult and the forces depend only on the
pairwise distances of the particles, the algorithmic complexity is
at least of the order $O(N^2)$ since each particle may interact with
each other. In fact
the probability that two particles which are a large
distance apart at time $t$, and interact within a small time interval $(t,
t+\Delta t)$ ($\Delta t$ is much larger than the integration step
$\delta t$) is very low, due to the probability distribution of
the particle's velocities. That means that even if there is a long range
force as the electrostatic force there is a critical threshold for the
distance beyond which the particles do not influence
each other. Therefore one can neglect the interaction of those
particles during some time interval $\Delta t$ and one can use
lists of neighbors $j$ for each
particle $i$ containing the particles $j$ which are closer to particle $i$
than a given threshold.\cite{verlet} The neighborhood--list has to be updated
each time interval $\Delta t$. The asymptotic time--complexity,
however, is not reduced because the calculation of the neighborhood
list is still of the order $O(N^2)$. Nevertheless the calculation time
reduces drastically.
Unfortunately in the general case there is no simple way to vectorize a
molecular dynamics code using neighborhood lists, e.g. Ref.~5.
There
are hierarchical force
calculation algorithms of
complexity $O(N\log(N))$
(Ref.~6), and even of complexity
$O(N)$
(Ref.~7). These algorithms make use of fast multipole expansions,
which cannot be applied to each type of particle interaction, as for the case
of short
range interaction, in particular if the force is not a steady
function of the distance between two interacting particles.
\par
The aim of this paper is to
describe a completely vectorizable algorithm for molecular dynamics, which
acts very effectively, provided
the following preconditions hold:
\par
\noindent
\begin{enumerate}
\item There is no long range interaction between the particles.
\item The particles are allowed to deform only slightly.
\item The dispersion of the particle size is not too large.
\item The particle density, i.e. the number of particles per space unit is
	high.
\end{enumerate}
\par
\noindent
These preconditions are provided in the molecular dynamics simulations of
granular material, as we will demonstrate below, which are of special interest.

\section{Description of the Algorithm}
Initially, we assume a region, where particles are allowed to move. One can
also
think about a periodically continued region to simulate periodic boundary
conditions. Now we define
a square lattice which covers this region and its cell size is determined by
the requirement that not more than one particle can reside per lattice cell.
Obviously the cell size $d$ is determined by the smallest particles and by
the
smallest distance $d_{min}$ they can have during the simulation $d=d_{min} /
\sqrt 2 $ (fig.~\ref{lattice}).
\par
\unitlength1.0cm
\begin{figure}[thb]
\caption{\it The size of the lattice cell is determined by the
smallest
particles and their minimum distance during the simulation since
there must not
be more than one particle residing within each lattice cell at time $t$.}
\label{lattice}
\vspace{2ex}
\end{figure}
\par
\noindent
Given this lattice our algorithm acts as follows:
\par
\noindent
\begin{enumerate}
\item Set the variables to their initial values.
\item Predict the positions and the time derivatives due to the
predictor--corrector algorithm as described in the Appendix.
\item Set up the lattice vectors:\\
\begin{eqnarray}
	\overline{Index}[i] & = & \left\{\begin{array}{cl}
		\mbox{particle
number} & \mbox{if there resides a particle}\\
		0 & \mbox{otherwise} \end{array} \right.\\[.2cm]
		  && i \in [1, LMAX] \nonumber\\ [.5cm]
	\overline{Test}[i] &  = & \left\{\begin{array}{cl}
		1 & \mbox{if there is a particle}\\
		0 & \mbox{otherwise} \end{array} \right.\\[.2cm]
                    && i \in [1, LMAX], \nonumber
\end{eqnarray}
where $LMAX$ is the number of lattice cells.
\par
The vector $\overline{Index}$ maps all observables which belong to the
particles to corresponding lattice vectors, so that we get a set of vectors
$\overline{x}$, $\overline{y}$, $\overline{F_x}$, $\overline{F_y}$, etc.

\item {Calculate the forces acting upon the particles.\\
Because the particles are assumed to have short range
interactions we only regard the interactions of a particle in cell $(i,j)$
with particles which reside
within cells $(k,l), (k\in [i-2,i+2], l\in[j-2,j+2], (k,l)\ne (i,j))$.
Therefore we may define a mask
(fig.~\ref{struct}) which describes the range of particle interactions. To
determine the interaction of each particle with the others we move the
centre
of the mask through all sites of the lattice and calculate the interactions of
the particles which might lie within the mask.
The index $i$, $i \in [1, 24]$, points to the mask positions which interacts
with
the regarded cell.
Hence the distances as well as all interesting values of the particles may be
expressed by 24 vectors indexed by the numbers of the lattice cells.
\par
\begin{equation}
\overline{dist}_i = \sqrt{\overline{\Delta x}_i \overline{\Delta x}_i +
\overline{\Delta y}_i \overline{\Delta y}_i} \hspace{1cm} i \in [1, 24]
\end{equation}
with
\begin{equation}
\overline{\Delta x}_i = \overline{x} - \overline{x\{i\}}
\end{equation}
\begin{equation}
\overline{\Delta y}_i = \overline{y} -\overline{y\{i\}},
\end{equation}
\par
\noindent
where the vector $\overline{x\{i\}}$ equals the
vector $\overline{x}$ shifted due to the position of $i$ according to the mask
enumeration as given in fig.~\ref{struct}.\\
\begin{figure}[thb]
\caption{\it Mask describing the area in the neighborhood of a particle, which
resides in lattice cell 0.}
\label{struct}
\vspace{2ex}
\end{figure}
\par
Now we define a set of vector variables $\overline{contact}_i$
$i\in [1,24]$ which checks whether or not the
particles touch each other.
\begin{equation}
\begin{array}{l}
\overline{contact}_i = \left\{\begin{array}{cl}
		1 & \mbox{if the particles touch each other}\\
		0 & \mbox{otherwise} \end{array} \right. \\
\hspace{2cm} i\in [1,24]. \end{array}
\end{equation}
With these variables the normal and shear forces read as:\\
\begin{equation}
\overline{F_n}_i = \overline{Test} \cdot \overline{Test\{i\}} \cdot
\overline{contact}_i
	\cdot {\cal F}_n(0,i)
\end{equation}
\begin{equation}
\overline{F_s}_i = \overline{Test} \cdot \overline{Test\{i\}} \cdot
\overline{contact}_i
	\cdot {\cal F}_s(0,i),
\end{equation}
where ${\cal F}_n(0,i)$ and ${\cal F}_s(0,i)$ denote the normal and shear
forces between particles which possibly reside in mask position 0 and $i$ ($i
\in [1, 24]$).\\
The total forces and the momentum acting upon each particle are given by:
\begin{equation}
\overline{F_x} = \sum_{i=1}^{24} \frac{1}{\overline{dist}_i}
\cdot (\overline{\Delta x}_i
	\cdot \overline{F_n}_i + \overline{\Delta y}_i \cdot \overline{F_s}_i)
\end{equation}
\begin{equation}
\overline{F_y} = \sum_{i=1}^{24} \frac{1}{\overline{dist}_i}
\cdot (\overline{\Delta y}_i
	\cdot \overline{F_n}_i - \overline{\Delta x}_i \cdot \overline{F_s}_i)
\end{equation}
\begin{equation}
\overline{M} = - \sum_{i=1}^{24} \overline{r} \cdot \overline{F_s}_i
\hspace{1cm}.
\end{equation}
}

\item Correct the predicted values due to the {\sc Gear}--algorithm as
described in the Appendix.
\item {Increase the time $t:=t + \delta t$.\\
	Proceed with step 2.}
\end{enumerate}
\par
\noindent
Assuming that the particles do not penetrate each other more than 10\% of
their radii, this algorithm works correctly if the ratio of the smallest and
the largest radii does not exceed $1:\frac{1.8}{\sqrt{2}}$. The cell size $d$
has to be set to the value of the maximum radius $(d=r_{max})$.\\
At each time--step one can extract all physical observables
of interest.\\

\section{Efficiency of the Algorithm}
To evaluate  the efficiency of the algorithm it is necessary to determine the
number of operations $NOP$ of each step described in Section 2 and the
corresponding vector length $n$. Thereby we disregard
the possibility of chaining. Given a pipeline depth $D$, one elementary
vector operation requires a time of $n+D$ clock periods instead of $n \cdot D$
which is the required time to process the equivalent code on a scalar machine.

For each part of the algorithm we get the results summarized in the following
table.\
\par
\vspace{3ex}
\begin{tabular}{|l|c|c|c|} \hline
Part of algorithm & $NOP$ & $n$ & estimated time\\ \hline
Predictor & 59 & $NP$ & $59 \cdot (NP+D)$\\ \hline
Index & 12 & $NP$ & $12 \cdot (NP+D)$\\ \hline
Forces & 1656 & $LMAX$ & $1656 \cdot (LMAX+D)$\\ \hline
Corrector & 54 & $NP$ & $ 54 \cdot (NP+D)$\\ \hline
\end{tabular}
\par
\noindent
\vspace{1ex}
$\begin{array}{ll}NP & \mbox{number of particles}\\
	LMAX & \mbox{number of lattice cells}
\end{array}$
\par
\noindent
One iteration step of the algorithm needs  the time:\\
$t=125 \cdot (NP+D)+1656 \cdot (LMAX+D)$ clock periods.\\
That means the time depends strongly on the number of lattice cells
but hardly on the number of particles.
Fig.~\ref{const} shows that the time of the vectorized algorithm does not
vary as rapidly with
increasing number of particles as the time of the scalar algorithm.
Figure~\ref{dens} shows the calculation time (normalized by the number of
particles) as a function of the number of particles for different particle
densities $\rho=\frac{NP}{LMAX}$. The required time for the
neighborhood--list method does not depend on the particle density,
because it works without the lattice.
\par
\begin{figure}[thb]
\caption{\it The processing time for vectorized calculations depends very
weakly
on the number of particles
while the time for the scalar neighborhood--list method rises as $N^2$. The
pipeline depth D was assumed to be 20.}
\label{const}
\vspace{2ex}
\end{figure}
\begin{figure}[thb]
\caption{\it Processing time as a function of the number of particles for
different particle densities compared to the neighborhood--list method (dashed
line).}
\label{dens}
\vspace{2ex}
\end{figure}
\indent
The density is obviously the limiting factor for the efficiency of the
algorithm. To be more effective than the neighborhood--list method,
the number of particles must increase with the reciprocal of the parameter
$\rho$.
Therefore the algorithm is constrained to problems with high particle density
or to very large systems with a lower density.\\

\section{MD-Simulation of granular material using the new algorithm}
The algorithm described above was intended to simulate the flow of granular
materials like dry sand. Moving sand reveals very astonishing effects. When
fluidized by shaking, it can behave like a fluid, while at rest it mostly
behaves like a
solid. Many experiments and
computer simulations of moving sand have been performed.\cite{sand}
\par
Using the proposed algorithm we investigated the flow of granular material
through a vertical
long narrow pipe with diameter $d$ and length $l$, ($l \gg d$) and
periodic boundary conditions.
We want to describe our numerical simulations here only as an
example to explain how the algorithm works and do not want to discuss the
numerical results in
detail which can be found in
Ref.~9.
To simulate a rough surface, the pipe was built of slightly smaller particles
which interact with the freely moving grains in the same manner as the freely
moving grains act on
each other
(fig.~\ref{pipe}).
\unitlength1.0cm
\begin{figure}[thb]
\caption{\it The narrow pipe has a diameter $d$ where only a few particles
are allowed to fit. Its surface consists of slightly smaller particles to
simulate a rough surface.}
\label{pipe}
\vspace{2ex}
\end{figure}
For this reason we need not distinguish between the interaction of the
grains with each other and between the grains and the wall. The particle radii
$R_i$
have been chosen from a {\sc Gauss}ian distribution with mean value $R_0$.
For the force we assumed the Ansatz given by Cundall and Strack\cite{cundall}
and slightly modified by Haff\cite{haff}:
\begin{equation}
	\vec{F}_{ij} =  \left\{ \begin{array}{cl}
		F_N \cdot \frac{\vec{r}_i - \vec{r}_j}{|\vec{r_i}-\vec{r}_j|}
		+ F_S \cdot  \left({0 \atop 1} ~{-1 \atop 0} \right) \cdot
                  \frac{\vec{r}_i - \vec{r}_j}{|\vec{r_i}-\vec{r}_j|}
			& \mbox{if $|\vec{r_i}-\vec{r}_j| < R_i + R_j$}
		\\[1ex]
		0
			& \mbox{otherwise}
		\end{array}
		\right.
\end{equation}
with
\begin{equation}
	F_N  = k_N \cdot (R_i + R_j - |\vec{r}_i - \vec{r}_j|)^{1.5}~+
	~\gamma_N \cdot M_{eff}\cdot (\dot{\vec{r}}_i -
\dot{\vec{r}}_j)
\end{equation}
and
\begin{equation}
	F_S = \min \{- \gamma_S \cdot M_{eff} \cdot v_{rel}~,~ \mu \cdot
	|F_N| \} ,
\label{eq_coulomb}
\end{equation}
where
\begin{equation}
	v_{rel} = (\dot{\vec{r}}_i - \dot{\vec{r}}_j) + R_i \cdot
\dot{\Omega}_i - R_j \cdot \dot{\Omega}_j)
\end{equation}
\begin{equation}
	M_{eff} = \frac{M_i \cdot M_j}{M_i + M_j} .
\end{equation}
\vspace{2ex}
\par
\noindent
The terms $\vec{r}_i$, $\dot{\vec{r}}_i$, $\dot{\Omega}_i$ and $M_i$ denote
the current position, velocity, angular velocity and mass of the $i$--th
particle. The model includes an elastic restoration force
which corresponds to the microscopic assumption that the particles can
slightly deform each other. In order
to mimic three--dimensional behaviour the  {\sc Hertz}ian
contact force\cite{landau} which increases with the power $\frac {3}{2}$
was applied. The other terms describe the energy dissipation of the system due
to collisions between particles according to normal and
shear friction. The parameters $\gamma_N$ and $\gamma_S$ stand for the normal
and shear friction coefficients. Eq.~\ref{eq_coulomb} takes into account that
the particles will not transfer rotational energy but slide on top of each
other if the relative velocity at the contact point exceeds a certain value
which depends on the normal component of the force acting between the particles
({\sc Coulomb} relation\cite{coulomb}).   For the parameters we have chosen
$\gamma_N=1000~s^{-1}$, $\gamma_S=300~s^{-1}$, $k_N=100~
N/m^{1.5}$ and $\mu=0.5$. The system showed accurate numerical behaviour when
the integration time step $\delta t = 10^{-4}~s$ was used.
\par
After starting the simulation with a homogeneous particle distribution, one
finds
the surprising result that the particles begin to form regions of different
density which can move in both directions. These density waves are of no
definite wavelength. The distances between the density maxima vary
irregularly with time
and depend strongly on the initial conditions. Figure~6 shows
the evolution of the pipe. The pipe is drawn every 500 time steps, and the
initial conditions are shown at the bottom.
After detecting the density waves in our numerical simulations they were
also found experimentally. For details see
Ref.~9.

\unitlength1.0cm
\begin{figure}[thb]
\caption{\it The simulated flow of granular  material through a pipe
is
shown in a sequence of snapshots. The snapshot on the bottom (t=0)
shows the
homogeneous particle distribution at the beginning of the simulation.
The pipe is plotted every 0.05 seconds starting at time t=10 sec.
The density profile varies irregularly with time and depends
essentially on the initial conditions. Gravity acts from left to
right.}
\label{plot}
\vspace{2ex}
\end{figure}

\section{Conclusion}
In conclusion we state that there is a lattice based algorithm of
complexity $O(N)$ which can be
vectorized completely and which acts very effectively given certain
preconditions, such as high particle density, low dispersion of the particle
size
and short range interactions between the particles. The algorithm was compared
in
detail with the neighborhood--list method. For the case of higher
particle dispersion, a similar algorithm can be defined using a mask of size $7
\times 7$ instead of $5
\times 5$. The algorithm was applied in a molecular dynamics simulation of
granular
material, and there were no differences in the results compared to equivalent
simulations using the classical neighborhood--list method.
\par
\nonumsection{Acknowledgments}
        The calculations were performed on a {\sc Cray-ymp el}
	vector
	computer.
        The authors thank the {\sc Max-Planck}-Institute for Nonlinear Dynamics
	at the University Potsdam for
	providing computer time and J.~Crepeau for a critical reading of the
	manuscript.

\nonumsection{References}


\begin{appendix}
\noindent
The algorithm of a {\sc Gear} predictor--corrector method reads as follows:
\par
\noindent
\begin{enumerate}
\item Predict the positions of the particles
$\vec{r}_i^{\hspace{0.03in}pr}(t+\delta t)$
	and the time derivatives
	$\dot{\vec{r}}_i^{\hspace{0.03in}pr}(t+\delta t)$,
	$\ddot{\vec{r}}_i^{\hspace{0.03in}pr}(t+\delta t)$,
	$\cdots$ up to the desired order of accuracy by
	{\sc Taylor} expansion
	using the known values at time $t$:
	\par
	\noindent
	$\vec{r}_i^{\hspace{0.03in}pr}(t+\delta t)=\vec{r}_i(t)
	+\delta t \cdot \dot{\vec{r}}_i(t)
	+\frac{1}{2} \delta t^2 \cdot \ddot{\vec{r}}_i(t)
	+\frac{1}{6} \delta t^3 \cdot \vec{r}_i^{\hspace{0.03in}(3)}(t)
	+\cdots$
	\nopagebreak
	\par
	\noindent
	$\dot{\vec{r}}_i^{\hspace{0.03in}pr}(t+\delta t)=\dot{\vec{r}}_i(t)
	+\delta t \cdot \ddot{\vec{r}}_i(t)
	+\frac{1}{2} \delta t^2 \cdot \vec{r}_i^{\hspace{0.03in}(3)}(t)
	+\cdots$
	\nopagebreak
	\par
	\noindent
	$\ddot{\vec{r}}_i^{\hspace{0.03in}pr}(t+\delta t)=\ddot{\vec{r}}_i(t)
	+\delta t \cdot \vec{r}_i^{\hspace{0.03in}(3)}(t)
	+\cdots$
	\nopagebreak
	\par
	\noindent
	\hskip1.5cm $\vdots$
\item Calculate the accelerations
	$\ddot{\vec{r}}_i^{\hspace{0.03in}corr}(t+\delta t)$ acting upon the
	particles using the predicted values.
\item Correct the predicted values using
	$\ddot{\vec{r}}_i^{\hspace{0.03in}corr}(t+\delta t)$ \hspace{0.05in}:
	\par
	\noindent
	$\begin{array}{cccccc}
	\left(\begin{array}{c}
	\vec{r}_i^{\hspace{0.03in}corr}(t+\delta t)\\
	\dot{\vec{r}}_i^{\hspace{0.03in}corr}(t+\delta t)\\
	\ddot{\vec{r}}_i^{\hspace{0.03in}corr}(t+\delta t)\\
	\vdots\end{array}\right) & = &
	\left(\begin{array}{c}
	\vec{r}_i^{\hspace{0.03in}pr}(t+\delta t)\\
	\dot{\vec{r}}_i^{\hspace{0.03in}pr}(t+\delta t)\\
	\ddot{\vec{r}}_i^{\hspace{0.03in}pr}(t+\delta t)\\
	\vdots\end{array}\right) & + &
	\left(\begin{array}{c}
	c_0\\
	c_1\\
	1\\
	\vdots\end{array}\right) &
	\ddot{\vec{r}}_i^{\hspace{0.03in}corr}(t+\delta t)
	\end{array}
	$
	\par
	\noindent
	where $c_i$ are constant values depending on the desired
	order of accuracy.\cite{predictor}
\item Proceed with the first step with incremented time $t:=t+\delta t$.

\end{enumerate}
\end{appendix}
\end{document}